\documentclass[10pt,onecolumn]{IEEEtran}
\usepackage{hyperref}
\usepackage{amsmath,amsthm,amssymb}
\usepackage{enumerate}
\usepackage{color}
\usepackage{url}
\usepackage{balance}
\usepackage{graphicx} 
\usepackage{mathrsfs}
\usepackage{epsfig}
\usepackage{verbatim}
\usepackage{setspace}
\usepackage{stfloats}

\newtheorem{theorem}{Theorem}
\newtheorem{lemma}[theorem]{Lemma}

\renewcommand{\vec}[1]{\mathbf{#1}}
\title{Dirty Paper Arbitrarily Varying Channel\\with a State-Aware Adversary}
\author{Amitalok~J.~Budkuley, Bikash~Kumar~Dey and Vinod~M.~Prabhakaran\\
   Emails: \{amitalok, bikash\}@ee.iitb.ac.in, vinodmp@tifr.res.in}
\begin{document}
\maketitle 
\thispagestyle{empty}
\begin{abstract}
In this paper, we take an arbitrarily varying channel (AVC) approach to examine the problem of writing on a dirty paper in the presence of an adversary. We consider an additive white Gaussian noise (AWGN) channel with an additive white Gaussian state, where the state is known non-causally to the encoder and the adversary, but not the decoder. We determine the randomized coding capacity of this  AVC under the maximal probability of error criterion. Interestingly, it is shown that the jamming adversary disregards the state knowledge to choose a white Gaussian channel input which is independent of the state. 
%
\end{abstract}
%
%
\section{Introduction}\label{sec:introduction}
In this paper, we study the problem of writing on a dirty paper in the presence of an adversary. In a celebrated paper~\cite{costa}, Costa determined the capacity of an AWGN channel with an additive white Gaussian state, where the state is known non-causally only to the encoder. Using the coding scheme of Gelfand and Pinsker~\cite{gelfand-pinsker}, he showed that the effect of the state can be completely nullified. The capacity of this ``dirty paper channel'' was shown to be equal to that of a standard AWGN channel with no state. Our aim in this work is to study communication over this dirty paper channel while under attack by a \emph{state-aware} jamming adversary. We model the communication channel as an \emph{arbitrarily varying channel} (AVC).

The AVC model was introduced by Blackwell et al. in a work~\cite{blackwell-ams1959} which studied the problem of communication over a channel where certain parameters of the channel were not known to the user and varied arbitrarily. The aim for the user was to communicate under any realization of these unknown parameters, which for instance, may be controlled by the adversary. It is observed that, in general, the results presented upon analysis of such AVC communication systems depend upon several factors, viz., possibility of randomization (unknown to the adversary) for the users, the probability of error criterion, assumptions on the jammer's knowledge, etc. 

Several works have analysed AVC models. Restricting the discussion to the continuous alphabet case (to which this paper primarily belongs), Hughes and Narayan in~\cite{hughes-narayan-it1987} analysed the Gaussian AVC and determined its capacity under the assumption of shared randomness and maximal probability of error criterion. Here, the jammer was assumed to be oblivious of the user's signal transmitted on the channel. The case where the jammer knew the transmitted codeword, was analysed in~\cite{agarwal}. In~\cite{sarwate-spcom2012}, Sarwate considered a `myopic' adversary, i.e., an adversary which listens to the channel and observes a noisy version of the transmitted codeword, and determined the capacity of such a Gaussian AVC. 
The scenario when the encoder possessed only private randomness but not randomness shared with the decoder, was analysed in~\cite{haddadpour-isit2013}. In a work related to ours, the Gaussian AVC with a state was studied in~\cite{sarwate-isit2008}. Unlike our setup, however, the adversary was assumed to be unaware of the random state. In addition, the user was allowed only deterministic codes while the error criterion was the average probability of error. 
For an in depth discussion on AVCs, one may refer~\cite{lapidoth-narayan-it1998},~\cite{sarwate-thesis} and the references therein. 

In this paper, we assume that the encoder and decoder share randomness which is unknown to the jammer and consider the maximal probability of error as the error criterion. Similar to the encoder, the adversary in this model is a \emph{state-aware} entity, i.e., it possesses a non-causal knowledge of the state. The main result of this work is the determination of the capacity of this Gaussian AVC. We show that the capacity achieving scheme is a dirty paper coding scheme. Interestingly, the state-aware adversary completely disregards the state knowledge and essentially performs independent and identically distributed (i.i.d.) Gaussian jamming, independent of the state.

This problem is related to our earlier work~\cite{amitalok-costa-isit2014}. There, for the very same AWGN channel with an additive white Gaussian state and a state-aware adversary, we took a game-theoretic approach to model the user-jammer interaction as a mutual information (zero sum) game~\cite{medard-1997} and analysed its Nash equilibrium. We defined the \emph{capacity} of the resulting channel as the \emph{unique} Nash equilibrium utility~\cite{owen} of this zero sum game, which we determined. We also identified an interesting elementary equilibrium pair of user and jammer strategies. We showed that at equilibrium, similar to the result in this work, the user chose a dirty paper coding scheme while the jammer performed i.i.d. Gaussian jamming, independent of state. 

The following is the organization of the paper. In Section~\ref{sec:system:model}, we describe the communication setup and provide the problem details. We state the main result of this work in Section~\ref{sec:main:result}. Next, in Section~\ref{sec:analysis} we perform the analysis and prove the main result. In Section~\ref{sec:discussion}, we briefly discuss the discrete memoryless channel version of this problem and make some overall concluding remarks in Section~\ref{sec:conclusion}.
\section{System Model and Problem Description}\label{sec:system:model}
\begin{figure}[!ht]
  \begin{center}
    \includegraphics[trim=0cm 10cm 0cm 1cm, scale=0.3]{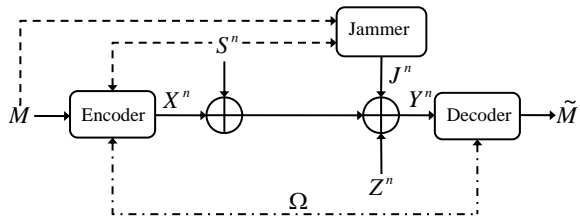}
    \caption{The Communication Setup}
    \label{fig:main:setup}
  \end{center}
\end{figure}
The dirty paper AVC setup is depicted in Fig.~\ref{fig:main:setup}. The transmitter aims to send a message $M$ to the receiver through $n$ channel uses in the presence of an adversary. The communication channel is an AWGN channel with an additive white Gaussian state and an additive jamming interference. The encoder and the decoder share an unbounded amount of common randomness, $\Omega$, unknown to the jammer. 
Let us denote by $\vec{Y}=(Y_1,Y_2,\ldots,Y_n)$, the signal received at the decoder. Then
\begin{equation*}\label{eq:Y}
\vec{Y}=\vec{X}+\vec{S}+\vec{J}+\vec{Z},
\end{equation*}
where $\vec{X}$, $\vec{S}$, $\vec{J}$ and $\vec{Z}$ are the encoder's input to
the channel, the additive white Gaussian state, jammer's channel input and the
channel noise respectively. The components of $\vec{S}$ are i.i.d. with
$S_i\sim\mathcal{N}(0,\sigma_S^2)$ for $i=1,2,\dots,n$. The components of
$\vec{Z}$ are i.i.d. with $Z_i\sim \mathcal{N}(0,\sigma^2)$ for $i=1,2,\dots,n$.
The state vector $\vec{S}$ is known non-causally to both the encoder and the
jammer, but it is not known to the decoder. The encoder picks a codeword
$\vec{X}=\vec{X}(M,\vec{S},\Omega)$ and transmits it on the channel.\footnote{We could
allow the transmitted codeword to depend additionally on a private randomness at
the encoder. But it is not difficult to see that such additional private
randomness at the encoder does not help the encoder-decoder pair in the presence
of unbounded common randomness.} The encoder has a power constraint $P$, i.e.
$\|\vec{X}\|^2 \leq n P$, where $\|.\|$ denotes the norm of a vector.
Similarly, the
adversary's power constraint is $\Lambda$ and hence, the signal $\vec{J}$ is
such that $\|\vec{J}\|^2 \leq n \Lambda$. Let $\mathcal{J}(\Lambda)=\left\{
\vec{J}: \|\vec{J}\|^2 \leq n \Lambda\right\}$. 
%

An $(n,R,P) $ \emph{deterministic code} of block length $n$, rate $R$ and
average power $P$ is a pair $(f,g)$ of encoder map
$f:\{1,2,\dots,2^{nR}\}\times \mathcal{\mathbb{R}}^n\rightarrow
\mathcal{\mathbb{R}}^n$, such that $\|f(m,\vec{s})\|^2\leq nP$ $\forall
m,\vec{s}$, and decoder map $g:
\mathcal{\mathbb{R}}^n\rightarrow \{1,2,\dots,2^{nR}\}$.

An $(n,R,P)$ \emph{randomized code} is a random variable $(F,G)$ ($=\Omega$ in our notation) which takes values in the set of $(n,R,P)$ deterministic codes.

For an $(n,R,P)$ randomized code with encoder-decoder pair $(F,G)$, 
the \emph{maximal probability of error} $(P^n_e)$ is given as
\begin{equation*}
P^n_e=\max_{m} \max_{p_{\vec{J}|m,\vec{S}}:\vec{J}\vec\in \mathcal{J}(\Lambda) } \mathbb{P}\left( G(F(m,\vec{S})+\vec{S}+\vec{J}+\vec{Z})\neq m\right) .
\end{equation*}
The rate $R$ is \emph{achievable} if for every $\epsilon>0$, there exists an
$(n,R,P)$ randomized code for some $n$ such that $P^n_e<\epsilon$ . We define
the \emph{capacity} of the dirty paper AVC as the supremum of all
achievable rates.
\section{The Main Result}\label{sec:main:result}
Our main contribution is the determination of the capacity of the Gaussian AVC with an additive white Gaussian state in the presence of a \emph{state-aware} adversary under a shared randomness and a maximal error probability criterion model. 
\begin{theorem}\label{thm:main:result}
The capacity of a Gaussian AVC with a Gaussian additive state in the presence of an adversary, where the encoder and the adversary have non-causal access to the state, is 
\begin{equation}\label{eq:capacity}
C=\frac{1}{2}\log \left( 1+\frac{P}{\Lambda+\sigma^2}\right).
\end{equation}
\end{theorem}
\noindent Note that this result implies that even under non-causal knowledge of the state $\vec{S}$, the adversary completely disregards this knowledge and essentially inputs i.i.d. Gaussian jamming noise.
\section{Proof of Theorem~\ref{thm:main:result}}\label{sec:analysis}
In this section, we discuss an achievable scheme for the main result stated in Section~\ref{sec:main:result}. Before we proceed, let us introduce some useful notation. For any $\vec{x}\in \mathbb{R}^n$, $\|\vec{x}\|\neq 0$, we denote the unit vector in the direction of $\vec{x}$ as $\hat{\vec{x}}$. Thus, $\hat{\vec{x}}=\vec{x}/\|\vec{x}\|$. Next, given two vectors $\vec{x},\vec{y} \in \mathbb{R}^n$, $\left<\vec{x},\vec{y}\right>\in \mathbb{R}$ denotes their dot (inner) product. 
\subsection{Codebook Construction}
Our code uses Costa's dirty paper coding scheme~\cite{costa}, which involves an auxiliary random variable denoted as $U$ and a fixed parameter $\alpha$. \\
\emph{Encoding:}
\begin{itemize}
\item  The encoder generates a code book comprising $2^{n R_U}=2^{n(R+\tilde{R})}$ i.i.d. vectors  $\{\vec{U}_{j,k}\}$, where $j=1,2,\dots,2^{nR}$ and $k=1,2,\dots,2^{n\tilde{R}}$. Here, there are $2^{nR}$ bins with each bin containing $2^{n\tilde{R}}$ codewords. Every codeword $\vec{U}_{j,k}$ is chosen uniformly at random over the surface of the $n$-sphere of radius $\sqrt{n P_U}$, where $P_U=P+\alpha^2 \sigma^2_S$ and $\alpha=P/(P+\Lambda+\sigma^2)$. 
\item Given a message $m$ to be sent and having observed a priori the state $\vec{S}$, the encoder looks within the bin $m$ for some $\vec{U}_{m,k}$, $k=1,2,\dots,2^{n\tilde{R}}$ such that 
\begin{equation}\label{eq:encoder:condition}
\|\left<\vec{U}_{m,k}-\alpha \vec{S},\vec{S}\right>\|\leq \delta_0
\end{equation}
for some appropriately small $\delta_0>0$. If no such $\vec{U}_{m,k}$ is found, then the encoder sends the zero vector. If more than one $\vec{U}_{m,k}$ satisfying~\eqref{eq:encoder:condition} exists, the encoder chooses one uniformly at random from amongst them. Let $\vec{U}=\vec{U}_{m,K}$ denote this codeword. The encoder then transmits $\vec{X}=\vec{U}-\alpha \vec{S}$ over the channel.  
\end{itemize}
\emph{Decoding:}\\
We employ the minimum angle decoder. When $\vec{Y}$ is received at the decoder, its estimate of the message, $\widetilde{m}$, is the solution of the following optimization problem.
\begin{equation*}
\widetilde{m}=\arg \max_{1\leq j\leq 2^{nR}}  \left(\max_{1\leq k\leq 2^{n\tilde{R}} } \left < \vec{\hat{Y}}, \vec{\hat{U}}_{j,k}\right>\right).
\end{equation*}
\subsection{Some Important Lemmas}
We now state some important results which are required toward the probability of error analysis of this code. The proof details for Lemmas~\ref{lem:binning:rate},~\ref{lem:J:U}, and~\ref{lem:Y:U:unit:expr} can be found in the appendix. Lemma~\ref{lem:csiszar:narayan} is recapitulated from~\cite{csiszar-narayan-it1991}. Finally, the proof of Lemma~\ref{lem:double:exp} is elementary, and hence, excluded. 

The following lemma gives a lower bound on $\tilde{R}$ (denoted by $\tilde{C}$) under which encoding succeeds with high probability.
\begin{lemma}[Binning Rate]\label{lem:binning:rate}
If $\tilde{R}>\tilde{C}=\frac{1}{2}\log\left(P_U/P\right)$, then the encoder finds at least one $\vec{U}_{m,k}$ satisfying~\eqref{eq:encoder:condition} with probability approaching 1 as $n\rightarrow \infty$.
\end{lemma}

The next result captures the correlation an adversary can induce with the codeword through the choice of its jamming signal. 
%
\begin{lemma}\label{lem:J:U}
For any $\delta>0$ and any jamming strategy $p_{\vec{J}|M,\vec{S}}:\vec{J}\in\mathcal{J}(\Lambda)$, 
\begin{equation*}
\mathbb{P}\left(\left| \left<\vec{J},\vec{U} \right>-\left<\vec{J},\vec{\hat{S}} \right>\left<\vec{\hat{S}},\vec{U} \right>\right| \geq n\delta \right)\rightarrow 0.
\end{equation*}
as $n\rightarrow\infty$.
\end{lemma}
An important result which directly follows from ~\cite[Lemma 2]{csiszar-narayan-it1991} is stated next. 
\begin{lemma}[\cite{csiszar-narayan-it1991}]\label{lem:csiszar:narayan}
Consider any $\vec{r}$ on the unit $n$-sphere and suppose an independent random vector $\vec{R}$ is uniformly distributed on this sphere. Then for any $1/\sqrt{2\pi n}<\gamma<1$, we have
\begin{equation*}\label{eq:CN}
\mathbb{P}\{ \left<\vec{r},\vec{R}\right>\geq \gamma\}\leq  2^{(n-1)\frac{1}{2}\log\left(1-\gamma^2\right)}, 
\end{equation*}
\end{lemma}
The following lemma shows that the inner product $\left<\vec{\hat{Y}},\vec{\hat{U}}\right>$ is at least $(\theta-\delta)$ with high probability irrespective of the jammer's strategy $p_{J|M,\vec{S}}:\vec{J}\in \mathcal{J}(\Lambda)$.
%
\begin{lemma}\label{lem:Y:U:unit:expr}
Let the codeword chosen be $\vec{U}$ and let $\vec{X}=\vec{U}-\alpha \vec{S}$ be transmitted on the channel which results in a channel output $\vec{Y}$. Then for any $\delta>0$ and any jamming strategy $p_{\vec{J}|M,\vec{S}}:\vec{J}\in\mathcal{J}(\Lambda)$, we have
\begin{equation*}\label{eq:tilde:Y:U:theta}
\mathbb{P}\left(\left<\vec{\hat{Y}},\vec{\hat{U}}\right><\left(\theta-\delta\right) \right)\rightarrow 0 
\end{equation*}
 as $n\rightarrow\infty$, where
\begin{equation}\label{eq:theta}
\theta=\sqrt{\frac{\alpha(P+\alpha \sigma_S^2)}{P_U  }}. 
\end{equation}
\end{lemma}

Finally, we close with the following result.
\begin{lemma}\label{lem:double:exp}
If $f(n)=2^{-na_1}$, where $a_1>0$, then
\begin{equation*}
\lim_{n\rightarrow \infty} (1-f(n))^{2^{na_2}}=
\begin{cases}
1 &\mbox{  if    } a_1>a_2\\
0 &\mbox{  if    } a_1<a_2
\end{cases}
\end{equation*}
\end{lemma}
\subsection{Probability of Error Analysis}
We start with a brief outline of the analysis. From Lemma~\ref{lem:Y:U:unit:expr} we know that regardless of the strategy the adversary employs,
a decoding error occurs only if any other codeword $\vec{U}_{m',k'}$, for some $m'\neq m$, $k'\in \{1,2,\dots, 2^{nR}\}$, is
such that $\left<\vec{\hat{Y}},\vec{\hat{U}}_{m',k'}\right>\geq (\theta-\delta)$. Our aim will be to show that this event has a vanishing probability.

\emph{Achievability:}
Fix some $\epsilon>0$, and let $R=C-\epsilon=\frac{1}{2}\log\left(1+P/(\Lambda+\sigma^2) \right)-\epsilon$ and $\tilde{R}=\tilde{C}+\epsilon/2=\frac{1}{2}\log(P_U/P)+\epsilon/2$. Hence,  $R_U=R+\tilde{R}=C+\tilde{C}-\epsilon/2=C_U-\epsilon/2$, where
\begin{equation*}
C_U=\frac{1}{2}\log\left(\frac{(P+\Lambda+\sigma^2)P_U}{(\Lambda+\sigma^2)P}      \right).
\end{equation*}

Let $\mathcal{E}_{m}$ denote the error event when $m$ is the message sent. Hence, we get
\begin{equation*}\label{eq:Em}
\mathcal{E}_{m}=\{\exists m'\neq m,\, k':\left<\vec{Y},\vec{U}_{m',k'}\right>\geq\left<\vec{Y},\vec{U}\right> \}.
\end{equation*}
Let $\theta$ be as in~\eqref{eq:theta}. For any $\delta>0$, conditioned on $M=m$
\begin{equation}\label{eq:P:Em:RHS:2} 
\mathbb{P}(\mathcal{E}_{m})\leq \mathbb{P}\left(\left<\vec{\hat{Y}},\vec{\hat{U}}\right><\theta-\delta\right)+\mathbb{P}\left(\exists m'\neq m,\, k': \left<\vec{\hat{Y}},\vec{\hat{U}}_{m',k'}\right> \geq \theta-\delta\Big| \left<\vec{\hat{Y}},\vec{\hat{U}}\right>\geq \theta-\delta\right).
\end{equation}
Lemma~\ref{lem:Y:U:unit:expr} implies that the first term can be made arbitrarily small. So, let us now consider the second term.
\begin{subequations}
\begin{IEEEeqnarray}{rCl}
\mathbb{P}\left(\exists m'\neq m,\,  k': \left<\vec{\hat{Y}},\vec{\hat{U}}_{m',k'}\right> \geq \theta-\delta \Big|\left<\vec{\hat{Y}},\vec{\hat{U}}\right>\geq\theta-\delta \right)&=&1-\mathbb{P}\left(\nexists m'\neq m,\, k': \left<\vec{\hat{Y}},\vec{\hat{U}}_{m',k'}\right>\geq\theta-\delta\Big| \left<\vec{\hat{Y}},\vec{\hat{U}}\right>\geq\theta-\delta \right)\nonumber\\
&\stackrel{(a)}{\leq}& 1- \left(1-\mathbb{P}\left(\left<\vec{\hat{Y}},\vec{\hat{U}}_{m',k'}\right>\geq \theta-\delta \Big| \left<\vec{\hat{Y}},\vec{\hat{U}}\right>\geq\theta-\delta\right)\right)^{2^{nR_U}}\\
&\stackrel{(b)}{\leq}& 1- \left(1-2^{(n-1)\frac{1}{2}\log\left(1-(\theta-\delta)^2\right)+1}\right)^{2^{nR_U}}\label{eq:P:e}
\end{IEEEeqnarray}
\end{subequations}
Due to the independence of vectors $\vec{U}$ and $\vec{U}_{m',k'}$, $m'\neq m$, $k'=\{1,2,\dots,2^{nR}\}$, we get $(a)$. To establish $(b)$, note firstly that $\vec{\hat{U}}_{m',k'}$ is independent of $(\vec{\hat{U}},\vec{\hat{Y}})$. In addition, recall Lemma~\ref{lem:csiszar:narayan}, and replace $(\vec{r},\vec{R})$ by $(\vec{\hat{Y}},\vec{\hat{U}}_{m',k'})$ and $\gamma$ by $(\theta-\delta)$. 
%
%

Now choosing a small enough $\delta>0\ $\footnote{Note that there exists $\delta>0$ such that~\eqref{eq:R_U:delta} is satisfied. To see this, define $h(\delta)=-1/2 \log(1-(\theta-\delta)^2)$. It can be easily verified that $h(0)=C_U$ and $h(.)$ is a continuous and a monotonically decreasing function of $\delta$.} such that 
\begin{equation}\label{eq:R_U:delta}
R_U<-\frac{1}{2}\log\left(1-(\theta-\delta)^2\right)
\end{equation}
and applying Lemma~\ref{lem:double:exp}, it can be seen that 
\begin{equation*}\label{eq:RHS2:tends:to:1}
\left(1-2^{(n-1)\frac{1}{2}\log\left(1-(\theta-\delta)^2\right)+1}\right)^{2^{nR_U}}\rightarrow 1.
\end{equation*}
as $n\rightarrow \infty$.
Thus, the second term of the RHS of~\eqref{eq:P:Em:RHS:2} can be made arbitrarily small. Thus, $P(\mathcal{E}_{m})$ can be made arbitrarily close to zero.
%

\emph{Converse:} Let the jammer choose a jamming signal $\vec{J}\perp\!\!\!\perp\vec{S}$ and $\vec{J}$ uniformly distributed over the sphere of radius $\sqrt{n\Lambda}$. We already know the capacity of such a channel and it is given by~\eqref{eq:capacity}.
%
%
\section{Discussion}\label{sec:discussion}
We now briefly discuss the discrete alphabet version of the AVC with a state-aware adversary. Let $\mathcal{X}$, $\mathcal{S}$, $\mathcal{J}$ and $\mathcal{Y}$ be finite alphabet sets. Let $x\in\mathcal{X}$ and $y\in\mathcal{Y}$ denote  resp., the user's input and the channel's output. We define $\mathcal{W}=\{W_{.|.,S,J}:S\in \mathcal{S},J\in \mathcal{J}\}$ as an AVC with a random state $S$ parametrized by an adversarial state $J$.

We use the notation $\mathcal{P}(\mathcal{A}|\mathcal{B})$ to denote the set of all conditional distributions $P_{A|B}$ for a random variable $A$ with alphabet $\mathcal{A}$ conditioned on a random variable $B$ with alphabet $\mathcal{B}$.
Next, let $\widetilde{\mathcal{W}}=\{V_{Y|X,S}: V_{Y|X,S}=\sum_{J} W_{Y|X,S,J} P_{J|S},  P_{J|S}\in \mathcal{P}(\mathcal{J}|\mathcal{S} )\}$.
Finally, given a state distribution $P_S$, for a fixed distribution $P_{U,X|S}$ and a fixed channel $V_{Y|X,S}\in\widetilde{\mathcal{W}}$, let $I(U;Y)$ and $I(U;S)$ denote resp., the mutual information quantities evaluated with respect to the marginals $P_{UY}$ and $P_{US}$. 

We now state without proof the following result.
\begin{theorem}[Capacity of Discrete Memoryless AVC]
The capacity of the discrete memoryless AVC with a random state, when both the encoder and the decoder have non-causal access to the state sequence is
\begin{equation*}
C=\max_{P_{U,X|S}\in \mathcal{P}(\mathcal{U}\times\mathcal{X}|\mathcal{S})} \min_{V\in \widetilde{W}} \big(I(U;Y)-I(U;S)\big).
\end{equation*}
\end{theorem}
\section{Conclusion}\label{sec:conclusion}
We determined the capacity of a Gaussian AVC with an additive Gaussian state in the presence of an adversary, where the state is known to the encoder as well as the adversary. The surprising fact that the worst-case adversary disregards state knowledge and inputs white Gaussian noise into the channel was proved. 
Overall,  it was shown that the effect of the state was completely eliminated and the capacity of a Gaussian AVC with state and a state-aware adversary is equal to that of a standard Gaussian AVC with no state and an independent adversary.
%
\section*{Acknowledgment}
The first author thanks Prof. Anand D. Sarwate for helpful early discussions and suggestions on the problem. 
\addcontentsline{toc}{section}{Acknowledgment} 

\bibliographystyle{IEEEtran}

\bibliography{IEEEabrv,References}

\begin{thebibliography}{10}
\providecommand{\url}[1]{#1}
\csname url@samestyle\endcsname
\providecommand{\newblock}{\relax}
\providecommand{\bibinfo}[2]{#2}
\providecommand{\BIBentrySTDinterwordspacing}{\spaceskip=0pt\relax}
\providecommand{\BIBentryALTinterwordstretchfactor}{4}
\providecommand{\BIBentryALTinterwordspacing}{\spaceskip=\fontdimen2\font plus
\BIBentryALTinterwordstretchfactor\fontdimen3\font minus
  \fontdimen4\font\relax}
\providecommand{\BIBforeignlanguage}[2]{{%
\expandafter\ifx\csname l@#1\endcsname\relax
\typeout{** WARNING: IEEEtran.bst: No hyphenation pattern has been}%
\typeout{** loaded for the language `#1'. Using the pattern for}%
\typeout{** the default language instead.}%
\else
\language=\csname l@#1\endcsname
\fi
#2}}
\providecommand{\BIBdecl}{\relax}
\BIBdecl

\bibitem{costa}
M.~M.~H. Costa, ``Writing on dirty paper (corresp.),'' \emph{IEEE Transactions
  on Information Theory}, vol.~29, pp. 439--441, 1983.

\bibitem{gelfand-pinsker}
S.~I. Gelfand and M.~Pinsker, ``Coding for channel with random parameters,''
  \emph{Problems of Control and Information Theory}, vol.~9, pp. 19--31, 1980.

\bibitem{blackwell-ams1959}
D.~Blackwell, L.~Breiman, and A.~J. Thomasian, ``The capacity of a class of
  channels,'' \emph{Annals of Mathematical Statistics}, vol.~30, no.~4, pp.
  1229--1241, 1959.

\bibitem{hughes-narayan-it1987}
B.~Hughes and P.~Narayan, ``Gaussian arbitrarily varying channels,'' \emph{IEEE
  Transactions on Information Theory}, vol.~33, pp. 267--284, 1987.

\bibitem{agarwal}
M.~Agarwal, A.~Sahai, and S.~Mitter, ``Coding into a source: a direct inverse
  rate-distortion theorem,'' \emph{Allerton Conference on Communication,
  Control and Computation}, 2006.

\bibitem{sarwate-spcom2012}
A.~Sarwate, ``An {AVC} perspective on correlated jamming,'' \emph{IEEE
  International Conference on Signal Processing and Communications}, 2012.

\bibitem{haddadpour-isit2013}
F.~Haddadpour, M.~Siavoshani, M.~Bakshi, and S.~Jaggi, ``On {AVC}s with
  quadratic constraints,'' \emph{IEEE International Symposium on Information
  Theory}, 2013.

\bibitem{sarwate-isit2008}
A.~Sarwate and M.~Gastpar, ``Arbitrarily dirty paper coding and applications,''
  \emph{IEEE International Symposium on Information Theory}, 2008.

\bibitem{lapidoth-narayan-it1998}
A.~Lapidoth and P.~Narayan, ``Reliable communication under channel
  uncertainty,'' \emph{IEEE Transactions on Information Theory}, vol.~44, pp.
  2148--2177, 1998.

\bibitem{sarwate-thesis}
A.~Sarwate, ``Robust and adaptive communication under uncertain interference,''
  Ph.D. dissertation, University of California, Berkeley, 2008.

\bibitem{amitalok-costa-isit2014}
A.~J. Budkuley, B.~K. Dey, and V.~M. Prabhakaran, ``Writing on dirty paper in
  the presence of jamming,'' \emph{International Symposium on Information
  Theory}, 2014.

\bibitem{medard-1997}
M.~M\'edard, ``Capacity of correlated jamming channels,'' \emph{Allerton Annual
  Conf. on Comm., Control and Computing}, 1997.

\bibitem{owen}
G.~Owen, \emph{Game Theory}.\hskip 1em plus 0.5em minus 0.4em\relax Emerald
  Group Publishing Limited, 1995.

\bibitem{csiszar-narayan-it1991}
I.~Csiszar and P.~Narayan, ``Capacity of the {G}aussian arbitrarily varying
  channel,'' \emph{IEEE Transactions on Information Theory}, vol.~37, pp.
  18--26, 1991.

\bibitem{cover-thomas}
T.~Cover and J.~Thomas, \emph{Elements of Information Theory}.\hskip 1em plus
  0.5em minus 0.4em\relax Wiley, New York, 1991.

\end{thebibliography}
\appendix
\subsection{Proof of Lemma~\ref{lem:binning:rate}}
\begin{proof}
From~\cite{costa}, we know that if $\tilde{R}>\frac{1}{n} I(\vec{U},\vec{S})$ then there exists at least one $\vec{U}_{m,k}$ satisfying~\eqref{eq:encoder:condition}. Recall that $\vec{U}=\vec{X}+\alpha \vec{S}$, where $\vec{X}$ and $\vec{S}$ are as described earlier. The result now follows by simply evaluating this mutual information quantity~\cite{cover-thomas}.
\end{proof}

\subsection{Proof of Lemma~\ref{lem:J:U}}
\begin{proof}
Let us resolve the components of $\vec{J}$ and $\vec{U}$ along directions parallel and orthogonal to $\vec{S}$. We denote the latter components as $\vec{J^{\perp}}$ and $\vec{U^{\perp}}$ respectively. 
\begin{IEEEeqnarray*}{rCl}\label{eq:J:U}
\vec{J}&=&\left<\vec{J},\vec{\hat{S}} \right>\vec{\hat{S}}+\vec{J^{\perp}}\nonumber\\
\vec{U}&=&\left<\vec{U},\vec{\hat{S}} \right>\vec{\hat{S}}+\vec{U^{\perp}}.\nonumber
\end{IEEEeqnarray*}
Note that $\left<\vec{J^{\perp}},\vec{\hat{S}} \right>=0=\left<\vec{U^{\perp}},\vec{\hat{S}} \right>$, and thus 
\begin{equation*}\label{eq:J:U:J:U:perp}
\left<\vec{J},\vec{U}\right>=\left<\vec{J},\vec{\hat{S}} \right>\left<\vec{\hat{S}},\vec{U} \right>+\left< \vec{J^{\perp}},\vec{U^{\perp}}\right>.
\end{equation*}
To establish the result in Lemma~\ref{lem:J:U} we need to show that for any $\delta>0$, as $n\rightarrow \infty$, $ \mathbb{P}\left(\lvert \left<\vec{J^{\perp}},\vec{U^{\perp}} \right>\rvert \geq n\delta \right)\rightarrow 0$, i.e., $\vec{J^{\perp}}$ and $\vec{U^{\perp}}$ are nearly orthogonal. To proceed, we introduce some notation. Let $\mathcal{S}^{n}\left(0,r\right)=\{\vec{w}\in \mathbb{R}^n:\|\vec{w}\|=r\}$ and for any $\vec{w}\in\mathbb{R}^n$, let $\mathcal{C}^{\perp}(\vec{w})$ denote the $(n-1)$ subspace orthogonal to $\vec{w}$.
\\
\indent\textit{\underline{Claim 1}}: Conditioned on $M=m$, $\vec{S}=\vec{s}$ and $\left<\vec{U},\vec{S}\right>= z$, the random vector $\vec{U}$ is uniformly distributed over  
\begin{equation}\label{eq:B}
\mathcal{B}_{z}(\vec{s})=\Big\{ z \frac{\vec{s}}{\|\vec{s}\|^2}+\vec{v}: \vec{v} \in \mathcal{S}^{n}\left(0,\rho_z(\vec{s})\right) \bigcap \mathcal{C}^{\perp}\left(\vec{s}\right)\Big\}.
\end{equation}
and, 
\begin{equation}\label{eq:rho:z:s}
\rho_z(\vec{s})=\sqrt{nP_U-\frac{z^2}{\|\vec{s}\|^2}}
\end{equation}
\indent\textit{\underline{Proof of Claim}}: Given the symmetry of the codebook generation and the encoding, we know that the chosen codeword vector $\vec{U}$ is uniformly distributed over the set $\mathcal{S}^n(0,\sqrt{n P_U})$. Now conditioned on message $M=m$, state $\vec{S}=\vec{s}$ and $\left<\vec{U},\vec{S}\right>=z$, it follows that the codeword vector $\vec{U}$ is uniformly distributed over the following set.
\begin{equation}\label{eq:B:tilde}
\mathcal{\tilde{B}}_{z}(\vec{s})=\big\{ \vec{u}: \|\vec{u}\|=\sqrt{n P_U} \text{   and  } \left<\vec{u},\vec{s}\right>=z\big\}.
\end{equation}
To proceed further, we show that $\mathcal{B}_z(\vec{s})=\tilde{\mathcal{B}}_z(\vec{s})$. 
The claim then follows from observing that $\vec{U}$ is uniformly distributed over the set $\mathcal{\tilde{B}}_{z}(\vec{s})$. 
\begin{enumerate}[i)]
\item To show $\vec{u}\in \mathcal{\tilde{B}}_{z}(\vec{s})\Rightarrow \vec{u}\in \mathcal{B}_{z}(\vec{s})$.\\
Let $\vec{u}\in \mathcal{\tilde{B}}_{z}(\vec{s})$. Expressing $\vec{u}$ through its two components, one in the direction parallel to $\vec{s}$ and the other orthogonal to it, we get
\begin{equation*}
\vec{u}=\left<\vec{u},\vec{s}\right>\frac{\vec{s}}{\|\vec{s}\|^2}+\vec{u}^{\perp}.
\end{equation*} 
Note here that $\left<\vec{u}^{\perp},\vec{s}\right>=0$ and $\|\vec{u}^{\perp}\|=\sqrt{n P_U-\frac{z^2}{\|\vec{s}\|^2}}$. Comparison with~\eqref{eq:B} completes the proof for the forward part.
\item To show $\vec{u}\in \mathcal{B}_{z}(\vec{s})\Rightarrow \vec{u}\in \mathcal{\tilde{B}}_{z}(\vec{s})$.\\
Consider some  vector $\vec{r}\in \mathcal{\tilde{B}}_z(\vec{s})$. Using~\eqref{eq:B:tilde}, we can write $\vec{r}=z\frac{\vec{s}}{\|\vec{s}\|^2}+\vec{v}$, $\vec{v}\in \mathcal{S}^{n}(0,\rho_z(\vec{s}))\bigcap \mathcal{C}^{\perp}\left(\vec{s}\right)$, and where $\rho_z(\vec{s})$ is as given in~\eqref{eq:rho:z:s}. It can be easily verified that $\|\vec{r}\|=\sqrt{n P_U}$. Also, $\left<\vec{v,\vec{s}}\right>=0$, and hence, it can be immediately seen that $\left<\vec{r},\vec{s}\right>=z$. Thus, $\vec{r}\in\mathcal{B}_z(\vec{s})$. 
\end{enumerate}

\indent\textit{\underline{Claim 2}}: For any $\delta>0$,  
\begin{equation*}
\mathbb{P}\left( \left| \left< \vec{J}^{\perp},\vec{U}^{\perp}\right> \right|> n\delta \right) \rightarrow 0.
\end{equation*}
\indent\textit{\underline{Proof of Claim}}: We first prove the conditional version of this claim. 
Again, let us condition on $M=m$, state $\vec{S}=\vec{s}$ and $\left<\vec{U},\vec{s}\right>=z$. From Claim $1$, we know that  $\vec{V}=\vec{U}(m,\vec{s})-\frac{z^2}{\|\vec{s}\|^2}$, 
$\vec{V}\sim \text{Unif} \left( \mathcal{S}^{n}(0,\sqrt{\rho_z(\vec{s})}) \bigcap C^{\perp}\left(\vec{s}\right)\right)$ with $\rho_z(\vec{s})$ as given in~\eqref{eq:rho:z:s}. Now for $\gamma>0$, we have
\begin{subequations}
\begin{IEEEeqnarray}{rCl}\label{eq:J:V:gamma}
\mathbb{P}\left( \frac{\left| \left< \vec{J}^{\perp},\vec{U}^{\perp}\right> \right|}{n} >\gamma \middle| M=m,\vec{S}=\vec{s}, \left<\vec{U},\vec{s}\right>=z  \right)&{=}&\mathbb{P}\left( \frac{1}{n}\left| \left< \frac{\vec{J}^{\perp}}{\|\vec{J}^{\perp} \|},\frac{\vec{V}}{{\|\vec{V}\|}}\right> \right| >\frac{\gamma}{{\|\vec{J}^{\perp} \| \|\vec{V}\|}} \middle| m,\vec{s}, z \right)\nonumber\\
&\stackrel{(a)}{\leq}&\mathbb{P}\left( \frac{1}{n}\left| \left< \vec{\hat{J}}^{\perp},\vec{\hat{V}}\right> \right| >\frac{\gamma}{ \sqrt{n\Lambda}\sqrt{nP_U} } \middle| m,\vec{s}, z \right)\\
&\stackrel{}{=}& \mathbb{P}\left( \left| \left< \vec{\hat{J}}^{\perp},\vec{\hat{V}}\right> \right| >\frac{\gamma}{\sqrt{ \Lambda P_U}}  \middle| m,\vec{s},z\right)\nonumber
\end{IEEEeqnarray}
\end{subequations}
Here, $(a)$ follows from noting that $\|\vec{J}^{\perp}\|\leq \|\vec{J}\|\leq \sqrt{n\Lambda}$  and $\|\vec{V}\|\leq \sqrt{nP_U}$. 

Since the shared randomness $\Omega$ is unavailable to the jammer, conditioned on $m$, $\vec{s}$ and $z$, we have $\vec{J}^{\perp}\perp\!\!\!\perp\vec{V}$. Also, both $\vec{J}^{\perp}$ and $\vec{V}$ lie in the $(n-1)$ hyperplane orthogonal to $\vec{s}$. Now using the result in Lemma~\ref{lem:csiszar:narayan}, we have
\begin{IEEEeqnarray*}{rCl}\label{eq:J:V:gamma'}
\mathbb{P}\left( \left| \left< \vec{\hat{J}}^{\perp},\vec{\hat{V}}\right> \right| >\gamma'  \middle| m,\vec{s},z\right)&\leq& 2^{-((n-1)-1)f(\gamma')}\nonumber\\
&=&2^{-(n-2)f(\gamma')} \,\,\,\forall \,m,\,\vec{s},\, z
\end{IEEEeqnarray*}
where $f(\gamma')=-\frac{1}{2}\log(1-\gamma'^2)=-\frac{1}{2}\log\left(P_U\Lambda/(P_U\Lambda-\gamma^2)\right)$ and $\gamma'=\gamma/\sqrt{\Lambda P_U}$. 
Since the upper bound in~\eqref{eq:J:V:gamma} tends to zero as $n\rightarrow \infty$, the conditional version of the result follows. However, note here that the bound in~\eqref{eq:J:V:gamma} does not depend on $m$, $\vec{s}$ or $z$. Hence, the unconditioned result is also true, and the claim follows.
\end{proof}
\subsection{Proof of Lemma~\ref{lem:Y:U:unit:expr}}
We know that
\begin{IEEEeqnarray}{rCl}\label{eq:Y:U:dot}
\left< \vec{Y},\vec{U}\right>&=&\left< \vec{U}+(1-\alpha)\vec{S}+\vec{J}+\vec{Z},\vec{U}\right>= \|\vec{U}\|^2+(1-\alpha)\left<\vec{S},\vec{U}\right>+\left<\vec{J},\vec{U} \right>+\left<\vec{Z},\vec{U} \right>
\end{IEEEeqnarray}
and
\begin{IEEEeqnarray}{rCl}
\|\vec{Y}\|^2	&=&\left< \vec{U}+(1-\alpha)\vec{S}+\vec{J}+\vec{Z},\vec{U}+(1-\alpha)\vec{S}+\vec{J}+\vec{Z}\right>\nonumber\\
&=& \|\vec{U}\|^2+(1-\alpha)^2\|\vec{S}\|^2+\|\vec{J}\|^2+\|\vec{Z}\|^2+2 ( \left<\vec{U},\vec{Z}\right>+\left<\vec{J},\vec{Z}\right>+\left<\vec{S},\vec{Z}\right>+(1-\alpha)\left( \left<\vec{U},\vec{S}\right>+\left<\vec{J},\vec{S}\right>\right)+\>\left<\vec{J},\vec{U}\right>)\label{eq:Y:Y}
\end{IEEEeqnarray}
Let $M=m$ and let us define the following events.
\begin{equation*}\label{eq:E:0}
E_0=\left\{\nexists\ k, \text{ s.t. } \left|\left<\vec{U}_{m,k}-\alpha\vec{S},\vec{S}\right>\right|\leq \delta_0 \right\}
\end{equation*}
\begin{equation*}\label{eq:E:1}
E_1=\left\{\left| \left<\vec{U,\vec{Z}}\right>\right|>n\delta_1\right\}
\end{equation*}
\begin{equation*}\label{eq:E:2}
E_2=\left\{\left| \left<\vec{S,\vec{Z}}\right>\right|>n\delta_2\right\}
\end{equation*}
\begin{equation*}\label{eq:E:3}
E_3=\left\{\left| \left<\vec{J,\vec{Z}}\right>\right|>n\delta_3\right\}
\end{equation*}
\begin{equation*}\label{eq:E:4}
E_4=\left\{\left|\|\vec{Z}\|^2-n\sigma^2\right|>n\delta_4\right\}
\end{equation*}
\begin{equation*}\label{eq:E:5}
E_5=\left\{\left|\|\vec{S}\|^2-n\sigma_S^2\right|>n\delta_5\right\}
\end{equation*}
\begin{equation*}\label{eq:E:6}
E_6=\left\{\left| \left<\vec{J,\vec{U}}\right>-\left<\vec{J,\vec{\hat{S}}}\right>\left<\vec{\hat{S}},\vec{U}\right>\right|>n\delta_6\right\}
\end{equation*}
%
%
%
From Lemma~\ref{lem:binning:rate}, $\mathbb{P}(E_0)\rightarrow 0$ as $n\rightarrow \infty$ for $\delta_0>0$. Since $\vec{Z}$ is independent of $\vec{U}$, $\vec{S}$ and $\vec{J}$, $\mathbb{P}(E_1)$, $\mathbb{P}(E_2)$ and $\mathbb{P}(E_3)\rightarrow 0$ as $n\rightarrow \infty$ for any $\delta_1>0$, $\delta_2>0$ and $\delta_3>0$ respectively. $\vec{Z}$ and $\vec{S}$ are i.i.d. Gaussian vectors with variance $\sigma^2$ and $\sigma_S^2$ resp., so for $\delta_4>0$, $\delta_5>0$, $\mathbb{P}(E_4)$, $\mathbb{P}(E_5)\rightarrow 0$ as $n\rightarrow \infty$. Finally, using Lemma~\ref{lem:J:U}, $\mathbb{P}(E_6)\rightarrow 0$ as $n\rightarrow\infty$ for any $\delta_6>0$.  

Let us define $E=\cup_{i=0}^6 E_i$ and let
\begin{align*}
V=\left<\vec{\hat{J}},\vec{\hat{S}}\right>\\
W=\frac{1}{n}\|J\|^2\label{eq:W:def}
\end{align*}
Since $\Big|\left<\vec{\hat{J}},\vec{\hat{S}}\right>\Big|\leq 1$, we have $V^2\leq 1$. It follows from $\|\vec{J}\|^2\leq n\Lambda$, that $0\leq W\leq \Lambda$. As argued above, $\mathbb{P}(E)\leq\epsilon$, where $\epsilon$ can be made arbitrarily small by choosing $n$ large enough for any $\delta_i$, $i=0,1,\dots,6$.
 
Recall that the codewords are chosen over the surface of an $n$-sphere of radius $\sqrt{n P_U}$, and hence, from~\eqref{eq:Y:U:dot},~\eqref{eq:Y:Y},  conditioned on the event $E^c$ 
\begin{IEEEeqnarray*}{rCl}
\left<\vec{Y},\vec{U}\right>&\geq&n\left(P_U+(1-\alpha)\alpha\sigma_S^2+V\alpha\sqrt{W\sigma_S^2}-\delta_a\right)\label{eq:Y:U:lb}
\end{IEEEeqnarray*}
and 
\begin{IEEEeqnarray*}{rCl}
\left<\vec{Y},\vec{Y}\right>&\leq& n\Big(P_U+(1-\alpha)^2\sigma_S^2+W+\sigma^2+2(1-\alpha)\alpha\sigma_S^2
+2V\alpha\sqrt{W\sigma_S^2}+2(1-\alpha)V\sqrt{W\sigma_S^2}+\delta_b\Big)\label{eq:Y:Y:ub}
\end{IEEEeqnarray*}
and thus, 
\begin{IEEEeqnarray*}{rCl}
\left<\vec{\hat{Y}},\vec{\hat{U}}\right>&\geq& \frac{\left(P_U+(1-\alpha)\alpha\sigma_S^2+V\alpha\sqrt{W\sigma_S^2}-\delta_a\right)}{\sqrt{P_U\Big(P_U+(1-\alpha)^2\sigma_S^2+W+\sigma^2+2(1-\alpha)\alpha\sigma_S^2+2V\alpha\sqrt{W\sigma_S^2}+2(1-\alpha)V\sqrt{W\sigma_S^2}+\delta_b\Big)  }}\nonumber\\
&\stackrel{(a)}{=}& \frac{\sqrt{\alpha}\left(P+\alpha \sigma_S^2+V \alpha \sqrt{W \sigma_S^2}-\delta_a\right)}{\sqrt{P_U\left(P+\alpha\sigma_S^2+\alpha(W-\Lambda)+2V\alpha\sqrt{W\sigma_S^2}+\alpha\delta_b\right)  }}\nonumber
\end{IEEEeqnarray*}
where, $\delta_a$, $\delta_b>0$ and $\delta_a$, $\delta_b\rightarrow 0$ as $\delta_i\rightarrow 0$, $i=0,1,\dots,6$.  Here, using $P_U=P+\alpha^2\sigma_S^2$ and $\alpha=P/(P+\Lambda+\sigma^2)$ and simplifying results in $(a)$. Furthermore, we have
\begin{IEEEeqnarray*}{rCl}
\left<\vec{\hat{Y}},\vec{\hat{U}}\right>&\geq& \frac{\sqrt{\alpha}\left(P+\alpha \sigma_S^2+V \alpha \sqrt{W \sigma_S^2}\right)}{\sqrt{P_U\left(P+\alpha\sigma_S^2+\alpha(W-\Lambda)+2V\alpha\sqrt{W\sigma_S^2}\right)  }}-\tilde{\delta}
\end{IEEEeqnarray*}
where, $\tilde{\delta}>0$ and $\tilde{\delta}\rightarrow0$ as $\delta_a$, $\delta_b\rightarrow 0$. 
%
%
Hence, it follows that $\mathbb{P}\left(\left<\vec{\hat{Y}},\vec{\hat{U}}\right>< (\theta-\delta)\right) \leq \mathbb{P}(E)\leq \epsilon$ if
\begin{IEEEeqnarray}{rCl}\label{eq:f:1}
\min_{V,W}\frac{\sqrt{\alpha}\left(P+\alpha \sigma_S^2+V \alpha \sqrt{W \sigma_S^2}\right)}{\sqrt{P_U\left(P+\alpha\sigma_S^2+\alpha(W-\Lambda)+2V\alpha\sqrt{W\sigma_S^2}\right)  }}-\tilde{\delta}>\theta-\delta
\IEEEeqnarraynumspace
\end{IEEEeqnarray}

The following claim completes the proof.
\\
\indent\emph{\underline{Claim}:} If 
\begin{IEEEeqnarray}{rCl}\label{eq:f:def}
f(V,W)=\frac{\sqrt{\alpha}\left(P+\alpha \sigma_S^2+V \alpha \sqrt{W \sigma_S^2}\right)}{\sqrt{P_U\left(P+\alpha\sigma_S^2+\alpha(W-\Lambda)+2V\alpha\sqrt{W\sigma_S^2}\right)}}
\end{IEEEeqnarray}
then, for all $-1\leq V\leq 1$ and $0\leq W\leq \Lambda$,  
\begin{equation*}\label{eq:f:geq:f:0:Lambda}
f(V,W)\geq \theta
\end{equation*}
where, $\theta=f(0,\Lambda)=\sqrt{\frac{\alpha(P+\alpha\sigma_S^2)}{P_U}}$.\\
\indent\emph{\underline{Proof of Claim}:} 
%
%
%
We show that for $-1\leq V\leq 1$  and $0\leq W\leq \Lambda$,
\begin{equation}\label{eq:f:fmin}
f(V,W)\geq f(0,\Lambda).
\end{equation}
Let us now establish the simple fact that $f(V,W)\geq 0$. Consider the numerator term in~\eqref{eq:f:def}.
\begin{subequations}
\begin{IEEEeqnarray}{rCl}
P+\alpha\sigma_S^2+V\alpha\sqrt{W\sigma_S^2}&= &P+\alpha\left( \sigma_S^2+V\sqrt{W\sigma_S^2}\right) \nonumber\\
&\stackrel{(a)}{=} &P+\frac{P}{P+\Lambda+\sigma^2}\left( \sigma_S^2+V\sqrt{W\sigma_S^2}\right) \IEEEyesnumber\\
&= &\frac{P}{P+\Lambda+\sigma^2}\left(P+\Lambda+\sigma^2+\sigma_S^2+V\sqrt{W\sigma_S^2}\right)\nonumber\\
&= &\frac{P}{P+\Lambda+\sigma^2}\left(P+\Lambda-W+W+\sigma^2+\sigma_S^2+V\sqrt{W\sigma_S^2}\right)\nonumber\\
&= &\frac{P}{P+\Lambda+\sigma^2}\left(P+\left(\Lambda-W\right)+\sigma^2+\left(W+\sigma_S^2+V\sqrt{W\sigma_S^2}\right)\right)\nonumber\\
&\stackrel{(b)}{\geq} &\frac{P}{P+\Lambda+\sigma^2}\left(P+\left(\Lambda-W\right)+\sigma^2+\left(W+\sigma_S^2-\sqrt{W\sigma_S^2}\right)\right)\IEEEyesnumber\\
&\geq &\frac{P}{P+\Lambda+\sigma^2}\left(P+\left(\Lambda-W\right)+\sigma^2+\left(W+\sigma_S^2-2\sqrt{W\sigma_S^2}\right)\right)\nonumber\\
&= &\frac{P}{P+\Lambda+\sigma^2}\left(P+\left(\Lambda-W\right)+\sigma^2+\left(\sqrt{W}-\sigma_S\right)^2\right)\nonumber\\
&\stackrel{(c)}{\geq} &0.
\end{IEEEeqnarray}
\end{subequations}
Here, $(a)$ follows by substituting for $\alpha$. Next, $(b)$ is true since $V>\geq-1$ while $(c)$ follows from noting that the parenthetic term in the step prior to $(c)$ is non-negative. Hence, we conclude that the numerator of~\eqref{eq:f:def} is non-negative, and $f(V,W)\geq 0$.
%

Now, since $f(V,\Lambda)\geq 0$ for $-1\leq V\leq 1$ and $0\leq W\leq\Lambda$, to show~\eqref{eq:f:fmin} it is sufficient to prove 
\begin{equation}\label{eq:f:squared}
(f(V,W))^2\geq (f(0,\Lambda))^2
\end{equation}
for $-1\leq V\leq1$ and $0\leq W\leq\Lambda$. Hence, by~\eqref{eq:theta} and~\eqref{eq:f:def}, we want to show that
\begin{IEEEeqnarray*}{rCl}
\left(\sqrt{\alpha}\frac{P+\alpha \sigma_S^2+V \alpha \sqrt{W \sigma_S^2}}{\sqrt{P_U(P+\alpha\sigma_S^2+\alpha(W-\Lambda)+2V\alpha\sqrt{W\sigma_S^2})  }}\right)^2 &{\geq}& \left(\sqrt{\alpha}\frac{\sqrt{P+\alpha \sigma_S^2}}{\sqrt{P_U  }}\right)^2\nonumber\\
\text{i.e.\,\,\,\,}\left(\frac{P+\alpha \sigma_S^2+V \alpha \sqrt{W \sigma_S^2}}{\sqrt{(P+\alpha\sigma_S^2+\alpha(W-\Lambda)+2V\alpha\sqrt{W\sigma_S^2})  }}\right)^2 &{\geq}& \left(\sqrt{P+\alpha \sigma_S^2}\right)^2\nonumber\\
\text{i.e.\,\,\,\,}\frac{\left(P+\alpha \sigma_S^2+V \alpha \sqrt{W \sigma_S^2}\right)^2}{(P+\alpha\sigma_S^2+\alpha(W-\Lambda)+2V\alpha\sqrt{W\sigma_S^2})  } &{\geq}& P+\alpha \sigma_S^2\nonumber\\
\text{i.e.\,\,\,\,}\left(P+\alpha \sigma_S^2+V \alpha \sqrt{W \sigma_S^2}\right)^2 &{\geq}& \left(P+\alpha \sigma_S^2\right)\left((P+\alpha\sigma_S^2+\alpha(W-\Lambda)+2V\alpha\sqrt{W\sigma_S^2} \right)\nonumber\\
\text{i.e.\,\,\,\,}\left(V \alpha \sqrt{W \sigma_S^2}\right)^2 &{\geq}& \left(P+\alpha \sigma_S^2\right)\alpha(W-\Lambda)
\end{IEEEeqnarray*}
Since $W\leq\Lambda$, the RHS above is negative. However, $-1\leq V\leq 1$, and hence, $V^2\geq 0$. Thus,~\eqref{eq:f:squared} immediately follows and we conclude that $f(V,W)\geq f(0,\Lambda)$, for $-1\leq V\leq 1$ and $W\leq\Lambda$. This concludes the proof of the claim.

Hence, using the result of the previous claim and from~\eqref{eq:f:1}, it can be seen that $\mathbb{P}\left(\left<\vec{\hat{Y}},\vec{\hat{U}}\right>< (\theta-\delta)\right)\leq \epsilon$ can be made arbitrarily small. 
This establishes the result in Lemma~\ref{lem:Y:U:unit:expr}.

\end{document}